\documentclass[twoside,a4paper,11pt]{sea10}
\usepackage{graphicx}
\usepackage{hyperref}
\usepackage{movie15}
\begin{document}
\pagenumbering{arabic}
\pagestyle{myheadings}
\thispagestyle{empty}
{\flushleft\includegraphics[width=\textwidth,bb=58 650 590 680]{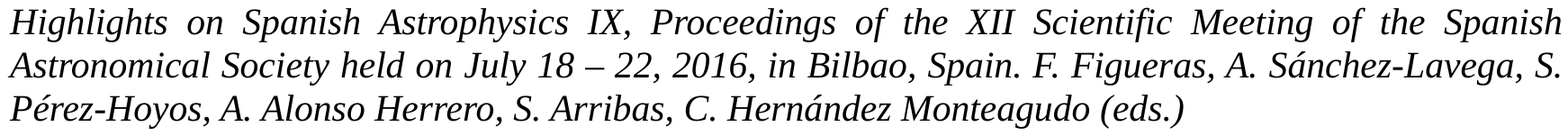}}
\vspace*{0.2cm}
\begin{flushleft}
{\bf {\LARGE
%
Study on the Abundance Discrepancy Problem in the Magellanic Clouds
%
}\\
\vspace*{1cm}
%
L. Toribio San Cipriano$^{1,2}$,
C. Esteban$^{1,2}$,
G.~Dom\'inguez-Guzm\'an$^{3}$,
and
J. Garc\'ia-Rojas$^{1,2}$
%
}\\
\vspace*{0.5cm}
%
$^{1}$
Instituto de Astrof\'isica de Canarias, E-38200, La Laguna, Tenerife, Spain\\
$^{2}$
Departamento de Astrof\'isica, Universidad de La Laguna, E-38206, La Laguna, Tenerife, Spain\\
$^{3}$
Instituto Nacional de Astrof\'isica, \'Optica y Electr\'onica INAOE, Apartado Postal 51, Puebla, Pue, Mexico\\

%
\end{flushleft}
%
\markboth{
Abundance Discrepancy Problem in MCs%
}{ 
%
L. Toribio San Cipriano et al.
%
}
\thispagestyle{empty}
\vspace*{0.4cm}
\begin{minipage}[l]{0.09\textwidth}
\ 
\end{minipage}
\begin{minipage}[r]{0.9\textwidth}
\vspace{1cm}
\section*{Abstract}{\small
%
We present chemical abundances of carbon (C) and oxygen (O) in the Large and Small Magellanic Clouds from  deep and high-quality optical spectra of H\thinspace\textsc{ii}\ regions. The data have been taken using the Ultraviolet-Visual Echelle Spectrograph at the 8.2-m Very Large Telescope  with the goal of detecting the faint C\thinspace\textsc{ii}\ and O\thinspace\textsc{ii}\ recombination lines. For all the objects of the sample, we determine $\mathrm{C^{2+}}$ abundances from recombination lines and $\mathrm{O^{2+}}$ abundances from recombination lines and collisionally excited lines. In addition, we calculate the abundance discrepancy factors (ADFs) for $\mathrm{O^{2+}}$ and $\mathrm{C^{2+}}$, as well as the O/H, C/H and C/O ratios. We study the behaviour of the ADF comparing the values obtained in the Magellanic Clouds with those obtained for other H\thinspace\textsc{ii}\ regions in different galaxies. We also compare the nebular and stellar abundances in two regions of the sample. Finally, we discuss the chemical evolution of the MCs through the O/H, C/H and C/O radial gradients and the changes of the C/O ratio as a function of O/H.
 
%
\normalsize}
\end{minipage}
%
%
%
\section{Introduction \label{intro}}

A proper knowledge of the chemical composition of extragalactic H\thinspace\textsc{ii}\ regions is essential for building chemical evolution models of galaxies. However, a well-known and unsolved issue in the physics of ionized nebulae is the so-called \textit{abundance discrepancy problem}. It refers to the difference between the chemical abundances determined from recombination lines (RLs) and from collisionally excited lines (CELs) of the same ion. Abundances based on RLs are systematically higher than those derived from CELs. The traditional method used to derive abundances is based on CELs because they are brighter and easier to detect than the optical RLs. On the other hand, the RLs are less dependent on the variations of the physical conditions and therefore, they provide more stable abundances.

Measurements of the faint RLs as C\thinspace\textsc{ii}\ 4267\,\AA\ and the multiplet 1 of O\thinspace\textsc{ii}\ at about 4650\,\AA\ have permitted us to calculate the $\mathrm{C^{2+}}$ and $\mathrm{O^{2+}}$ abundances for a sample of H\thinspace\textsc{ii}\ regions in the Large Magellanic Cloud (LMC) and the Small Magellanic Cloud (SMC). For each H\thinspace\textsc{ii}\ regions, we computed the abundance discrepancy factor (ADF), which is defined as:
\begin{equation}
\mathrm{ADF(X^{i+}) \equiv log(X^{i+}/H^+)_{RLs} - log(X^{i+}/H^+)_{CELs}}
\end{equation}
where $\mathrm{X^{i+}}$ corresponds to the ionization state i of the element X. 

\section{The sample}
The sample comprises 5 H\thinspace\textsc{ii}\ regions in the LMC and 4 in the SMC. We observed 7 of these objects using the Ultraviolet Visual Echelle Spectrograph (UVES) at the Very Large Telescope (VLT) at Cerro de Paranal Observatory (Chile) on 2003 March and 2013 November. The data of the other two H\thinspace\textsc{ii} regions were taken from \cite{2003ApJ...584..735P} and \cite{2012ApJ...746..115P}. In order to have an homogeneous data set, in these two cases, we took the measured line fluxes and performed the analysis in the same way than for the rest of the sample. 

In this work we also used the C and O abundances of H\thinspace\textsc{ii} regions in other galaxies taken from the literature: the Milky Way (\cite{2007ApJ...670..457G, 2006MNRAS.368..253G, 2005MNRAS.362..301G, 2004ApJS..153..501G, 2004MNRAS.355..229E, 2013MNRAS.433..382E}), M101 (\cite{2009ApJ...700..654E}), M33 (\cite{2016MNRAS.458.1866T}) and dwarf galaxies (\cite{2014MNRAS.443..624E}).

\section{Nebular abundances and metallicity-dependence of the ADF}

We calculated the ADF($\mathrm{O^{2+}}$) for all the objects of the sample. The values show a fairly low dispersion in both galaxies. Only the N44C region, which has He\thinspace\textsc{ii} lines in its spectrum, presents an ADF($\mathrm{O^{2+}}$)
higher than the rest of the objects of the LMC. We also computed the ADF($\mathrm{C^{2+}}$) using the $\mathrm{C^{2+}}$ abundances derived from CELs in the UV by \cite{1995ApJ...443...64G} and \cite{1982ApJ...252..461D}. The results show a higher dispersion than the ADF($\mathrm{O^{2+}}$). Table \ref{tab:adf} contains the ADF of $\mathrm{O^{2+}}$ and $\mathrm{C^{2+}}$ for the H\thinspace\textsc{ii} regions in the LMC and SMC.

\begin{table*}
   \centering
   \caption{ADF of $\mathrm{O^{2+}}$ and $\mathrm{C^{2+}}$ for H\thinspace\textsc{ii} regions in the LMC and SMC.}
   \label{tab:adf}
    \medskip
   \begin{tabular}{lccccc}
        \hline \hline
         \multicolumn{6}{c}{LMC}\\
                               &  30 Doradus     & N44C           &    IC2111      &     NGC1714    &  N11B           \\

        ADF(O$^{2+}$)          &$0.14 \pm 0.05$  &$0.31 \pm 0.03$ &$0.18 \pm 0.12$ &$0.20 \pm 0.09$ &$0.20 \pm 0.04$ \\
        ADF(C$^{2+}$)          &$0.22 \pm 0.23$  &--              &$0.01 \pm 0.25$ &$0.10 \pm 0.25$ &       --       \\

        \hline
   \end{tabular}
   \begin{tabular}{lcccc}
        \multicolumn{5}{c}{SMC}\\
                               &  N66A           &    N81         &   NGC456       &   N88A              \\

        ADF(O$^{2+}$)           &$0.35 \pm 0.13$ &$0.33 \pm 0.11$ &$0.29 \pm 0.13$ &$0.29 \pm 0.10$      \\

        ADF(C$^{2+}$)          &$0.45 \pm 0.10$  &$0.43 \pm  0.10$ &    --          &$0.38 \pm 0.04$      \\

        \hline
   \end{tabular}

\end{table*}

\begin{figure*}
 \centering
  \includegraphics[width=0.49\textwidth]{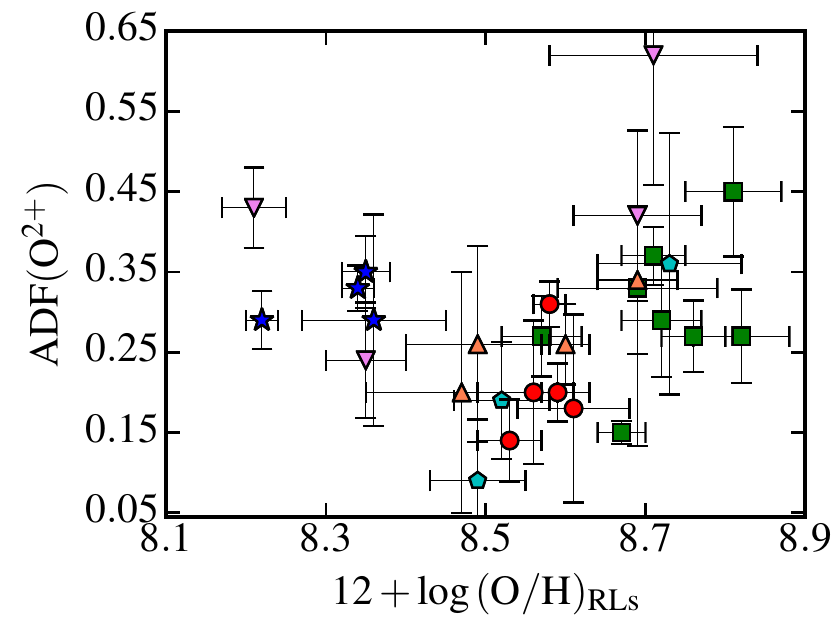}
  \includegraphics[width=0.49\textwidth]{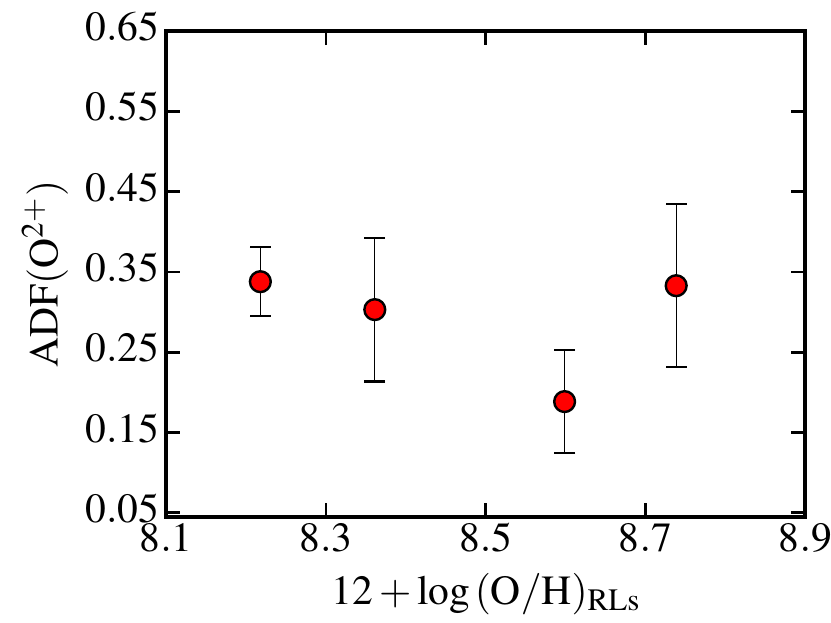}
 \caption{ADF(O$^{2+}$) {\it versus} O/H ratio derived from RLs. In the left-hand panel we show data for H\thinspace\textsc{ii}\ regions in different galaxies. The right-hand panel represents the average of ADF(O$^{2+}$) for different abundance bins.}
 \label{fig:F1}
\end{figure*}

In order to explore a possible metallicity-dependence of the ADF, we studied the behaviour of the ADF($\mathrm{O^{2+}}$) in the H\thinspace\textsc{ii}\ regions of several galaxies as a function of the total O abundance. 
The left-panel of Fig.~\ref{fig:F1} shows the results when we use the O abundances based in RLs.  We computed the average of ADF($\mathrm{O^{2+}}$) for different metallicity bins of the O/H ratio in the right panel of Fig.~\ref{fig:F1}. The results are difficult to interpret. The ADF($\mathrm{O^{2+}}$) shows a minimum value at 12 + log(O/H) $\sim$ 8.5 and increase when the O abundances are higher and lower drawing a seagull shape. The remarkably consistent value of the ADF($\mathrm{O^{2+}}$) obtained for the H\thinspace\textsc{ii}\ regions of each one of the MCs has been the main reason to search for a metallicity-dependent behaviour of the ADF. The trend we found for the low-metallicity objects seems to be clear, but the high dispersion at the highest metallicities (12 + log O/H $>$ 8.7) has blurred any clear trend.

than the high-metallicity ones due to the high dispersion of the ADF in this last brach of metallicity. 

We also studied the ADF($\mathrm{C^{2+}}$) as a function of the total O abundances. We find an apparent trend of a slightly larger ADF($\mathrm{C^{2+}}$) at lower abundances. Unfortunately, there are only two determinations of the ADF(C$^{2+}$) for metal-rich Galactic H\thinspace\textsc{ii}\ regions and they are uncertain, given the large aperture effects in extended H\thinspace\textsc{ii}\  regions. However, their ADF(C$^{2+}$) are also larger than those of the objects with 12 + log(O/H) $\sim$ 8.5, in agreement with the general behaviour shown in Fig.~\ref{fig:F1}.

\section{Comparison between nebular and stellar abundances}

An interesting test to explore the abundance discrepancy  problem is to compare the abundances derived from CELs and RLs in H\thinspace\textsc{ii}\ regions with those determined in B-type stars located in their vicinity. The O abundance of B-type stars should reflect the present-day chemical composition of the interstellar material in the regions where they were formed and are located.

\begin{figure*}
 \centering
  \includegraphics[width=0.49\textwidth]{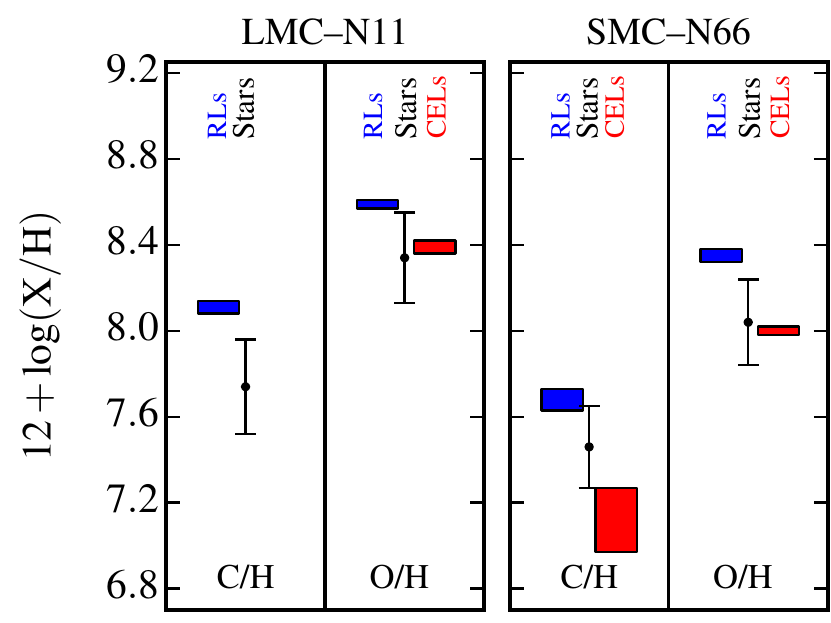}
 \caption{Comparison between C and O abundances of two  H\thinspace\textsc{ii}\ regions obtained from CELs (red rectangles) and RLs (blue rectangles) and B-type supergiant stars belonging to the associated clusters \cite{2009A&A...496..841H} (black dots and error bars). }
 \label{fig:F2}
\end{figure*}

In the left-panel of Fig.~\ref{fig:F2} we compared our C and O nebular abundances with those determined by \cite{2009A&A...496..841H} for B-type stars in the H\thinspace\textsc{ii}\ regions N11 in the LMC and N66 in SMC. We conclude that O abundances derived for B-type stars agree better with the nebular ones derived from CELs than with those derived from RLs. The C abundances determined from RLs are too high and inconsistent with the stellar ones when we assume that part of the C is embedded in dust grains (0.10\,dex see \cite{1998MNRAS.295..401E}).

\section{The spatial distribution of O and C abundances}

We studied the spatial distribution of O and C abundances as a function of the fractional galactocentric distances ($R/R_{25}$) in the LMC and SMC. Table~\ref{tab:fits} shows the radial abundance gradients for both galaxies. We find that the slope of the O/H gradient is almost the same using CELs or RLs and can be considered essentially flat in both galaxies.  Moreover, the slope of  C/H and C/O gradients in both galaxies can also be considered practically zero, in contrast with the results that are found in spiral galaxies (e.g. \cite{2009ApJ...700..654E, 2016MNRAS.458.1866T}).

\begin{table}
     \caption{Radial abundance gradients for the LMC and SMC.}
     \center
     \begin{minipage}{0.5\textwidth}
     \center
     \label{tab:fits}
     \begin{tabular}{@{}lccc}
           \hline
           \hline
                                        & Lines  &\multicolumn{2}{c}{slope$\big( \mathrm{dex}\  (R/R_{25})^{-1} \big)$}            \\
           \hline
                                        &        &  LMC              &  SMC             \\
          
           $12 +  \log(\mathrm{O/H})$   & CELs   &  $0.05 \pm 0.05$  & $-0.03 \pm 0.03$  \\
                                        & RLs    &  $0.04 \pm 0.06$  & $-0.08 \pm 0.04$  \\
           $12 + \log(\mathrm{C/H})$    & RLs    &  $0.05 \pm 0.05$  & $0.01 \pm 0.06$   \\ 
           $ \log(\mathrm{C/O})$        & RLs    &  $0.01 \pm 0.08$  & $0.09 \pm 0.07$   \\ 
           \hline     
 \end{tabular}
 \end{minipage}
 \medskip 
\end{table}
\begin{figure*}
 \centering
  \includegraphics[width=0.49\textwidth]{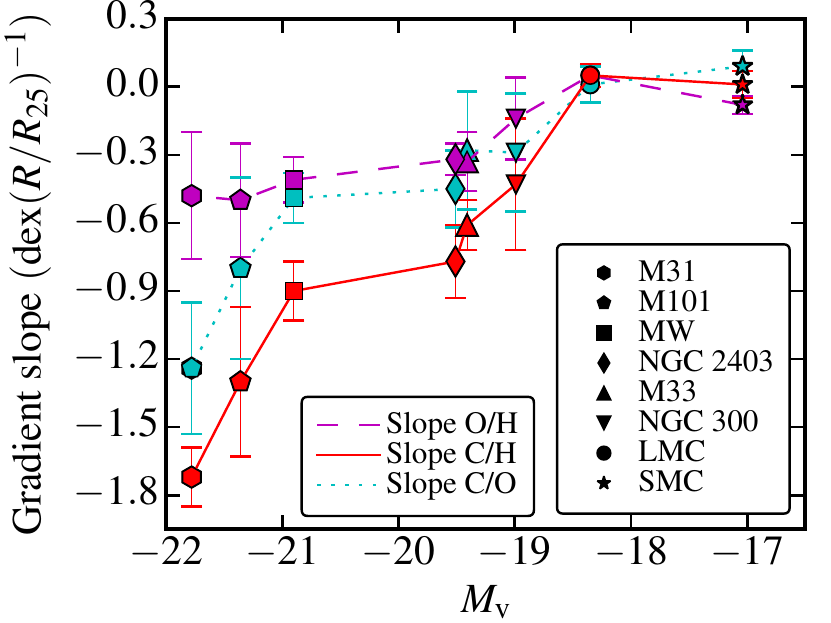}
  \includegraphics[width=0.49\textwidth]{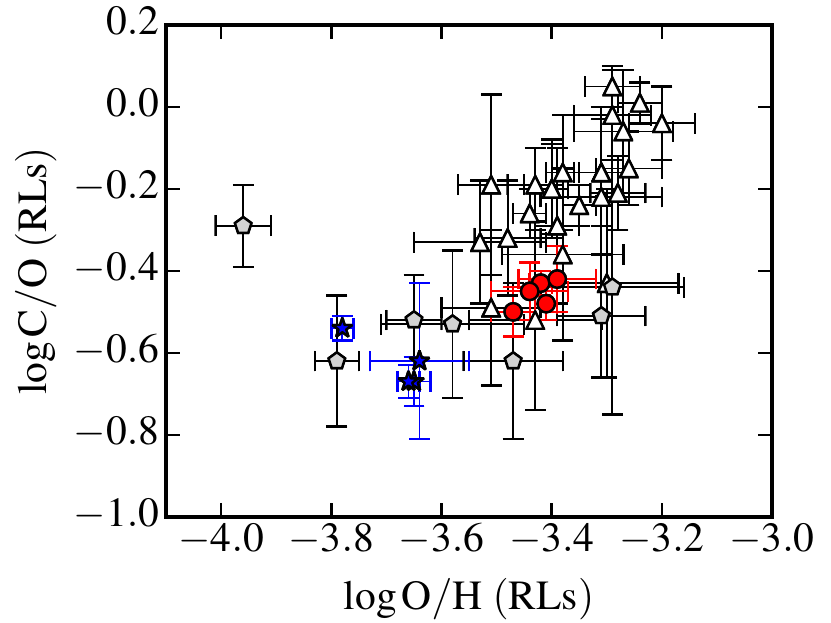}
 \caption{\textit{Left-panel}: Slope of O/H (magenta dashed line/symbols), C/H (red continuous line/symbols) and C/O (cyan dotted line/symbols) radial gradients versus absolute magnitude ($M_\mathrm{V}$) for several galaxies. \textit{Right-panel}: C/O \textit{versus} O/H ratios of H\thinspace\textsc{ii}\ regions in different galaxies: LMC (red circles), SMC (blue stars), star-forming dwarf galaxies (gray pentagons) and other spiral galaxies (open triangles).}
 \label{fig:F3}
\end{figure*}

The left-panel of Fig.~\ref{fig:F3} represents the slope of O/H, C/H and C/O gradients for different nearby spiral galaxies. The results we found strengthen those found by \cite{2016MNRAS.458.1866T}, where C/H and C/O gradients show a clear correlation with the absolute magnitude $M_\mathrm{V}$ of the galaxies. The more luminous galaxies show steeper slopes of C and C/O than the less luminous ones.

In addition, we studied the behaviour of C/O versus O/H ratios determined from RLs in the right-panel of Fig.~\ref{fig:F3}. The position of the H\thinspace\textsc{ii}\ regions in the figure suggests that the chemical evolution of the SMC behaves as a typical star-forming dwarf galaxy, while the LMC seems to be similar to the external zones of small spiral galaxies.

%
%
\small  
%
\section*{Acknowledgments}   
%
This work is based on observations collected at the European
Southern Observatory, Chile, proposal numbers ESO 092.C-0191(A) and ESO 60.A-9022(A). LTSC is supported by the FPI
Program by the Ministerio de Econom\'{i}a y Competitividad
(MINECO) under grant AYA201122614. This project project is also partially funded by MINECO under grant AYA2015-65205-P.

%

%
\end{document}